\documentstyle[12pt]{article}
\newcommand{\n}{\hspace*{-2.5mm}}
\newcommand{\gsim}{\;\rlap{\lower 3.5 pt \hbox{$\mathchar \sim$}} \raise 1pt
 \hbox {$>$}\;}
\newcommand{\lsim}{\;\rlap{\lower 3.5 pt \hbox{$\mathchar \sim$}} \raise 1pt
 \hbox {$<$}\;}
\newcommand{\di}{\mathop{\mbox{Li}_2}\nolimits}

\catcode`@=11
\newcount\@tempcntc
\def\@citex[#1]#2{\if@filesw\immediate\write\@auxout{\string\citation{#2}}\fi
  \@tempcnta\z@\@tempcntb\m@ne\def\@citea{}\@cite{\@for\@citeb:=#2\do
    {\@ifundefined
       {b@\@citeb}{\@citeo\@tempcntb\m@ne\@citea\def\@citea{,}{\bf ?}\@warning
       {Citation `\@citeb' on page \thepage \space undefined}}%
    {\setbox\z@\hbox{\global\@tempcntc0\csname b@\@citeb\endcsname\relax}%
     \ifnum\@tempcntc=\z@ \@citeo\@tempcntb\m@ne
       \@citea\def\@citea{,}\hbox{\csname b@\@citeb\endcsname}%
     \else
      \advance\@tempcntb\@ne
      \ifnum\@tempcntb=\@tempcntc
      \else\advance\@tempcntb\m@ne\@citeo
      \@tempcnta\@tempcntc\@tempcntb\@tempcntc\fi\fi}}\@citeo}{#1}}
\def\@citeo{\ifnum\@tempcnta>\@tempcntb\else\@citea\def\@citea{,}%
  \ifnum\@tempcnta=\@tempcntb\the\@tempcnta\else
   {\advance\@tempcnta\@ne\ifnum\@tempcnta=\@tempcntb \else \def\@citea{--}\fi
    \advance\@tempcnta\m@ne\the\@tempcnta\@citea\the\@tempcntb}\fi\fi}
\catcode`@=12
\begin{document}
\title{\vskip-3cm{\baselineskip14pt
\centerline{\normalsize\hfill DOE/ER/40561--226--INT95--17--11}
\centerline{\normalsize\hfill MPI/PhT/95--84}
\centerline{\normalsize\hfill hep-ph/9602304}
\centerline{\normalsize\hfill August 1995}
}
\vskip1.5cm
Two-Loop ${\cal O}(\alpha_sG_FM_Q^2)$ Heavy-Quark Corrections to the
Interactions between Higgs and Intermediate Bosons}
\author{{\sc Bernd A. Kniehl}\thanks{Permanent address:
Max-Planck-Institut f\"ur Physik, Werner-Heisenberg-Institut,
F\"ohringer Ring 6, 80805 Munich, Germany.}\\
{\normalsize National Institute for Nuclear Theory, University of Washington,}
\\
{\normalsize P.O. Box 351550, Seattle, WA 98195, USA}}
\date{}
\maketitle
\begin{abstract}
By means of a low-energy theorem, we analyze at ${\cal O}(\alpha_sG_FM_Q^2)$
the shifts in the Standard-Model $W^+W^-H$ and $ZZH$ couplings induced by
virtual high-mass quarks, $Q$, with $M_Q\gg M_Z,M_H$, which includes the top
quark.
Invoking the improved Born approximation, we then find the corresponding
corrections to various four- and five-point Higgs-boson production and decay
processes which involve the $W^+W^-H$ and $ZZH$ vertices with one or both of
the gauge bosons being connected to light-fermion currents, respectively.
This includes $e^+e^-\to f\bar fH$ via Higgs-strahlung, via $W^+W^-$
fusion (with $f=\nu_e$), and via $ZZ$ fusion (with $f=e$),
as well as $H\to2V\to4f$ (with $V=W,Z$).

\end{abstract}

\section{Introduction}

The Higgs boson is the last missing link in the Standard Model (SM).
The discovery of this particle and the study of its characteristics are among
the prime objectives of present and future high-energy colliding-beam
experiments.
Following Bjorken's proposal \cite{bjo}, the Higgs boson is currently being
searched for with the CERN Large Electron-Positron Collider (LEP1) and the
SLAC Linear Collider (SLC) via $e^+e^-\to Z\to f\bar fH$.
At the present time, the failure of this search allows one to rule out the
mass range $M_H\le64.3$~GeV at the 95\% confidence level \cite{jan}.
The quest for the Higgs boson will be continued with LEP2 by exploiting the
Higgs-strahlung mechanism \cite{ell,iof}, $e^+e^-\to ZH\to f\bar fH$.
In next-generation $e^+e^-$ linear supercolliders (NLC), also
$e^+e^-\to \nu_e\bar\nu_eH$ via $W^+W^-$ fusion and, to a lesser extent,
$e^+e^-\to e^+e^-H$ via $ZZ$ fusion will provide copious sources of Higgs
bosons.

The study of quantum corrections to the production and decay processes of the
Higgs boson has received much attention in the literature; for a review,
see Ref.~\cite{kni}.
Since the top quark, with pole mass $M_t=(180\pm12)$~GeV \cite{abe}, is so much
heavier than the intermediate bosons, the $M_t$-dependent corrections are
particularly important.
On the other hand, it is attractive to consider the extension of the SM by
a fourth fermion generation, where such corrections may be even more 
significant.
Some time ago, Hill and Paschos \cite{hil} proposed an interesting
fourth-generation scenario with Majorana neutrinos, which exploits the
see-saw mechanism to evade the LEP1/SLC constraint on the number of light
neutrinos.
The novel charged fermions of this model are assumed to be of Dirac type and to 
have standard couplings.
Subsequently, this model was further elaborated, and the precise triviality
bounds, renormalization-group fixed points, and related dynamical mechanisms
were discussed \cite{lut}.
In particular, it was demonstrated how this model is reconciled with the
fermion-mass constraints established in Ref.~\cite{cab}.
In Ref.~\cite{ber}, it was shown that this model is compatible with precision
data from low energies and LEP1/SLC.
Very recently, it was noticed \cite{cel} that arguments favouring the presence
of a fourth fermion generation may be adduced on the basis of the democratic
mass-matrix approach \cite{fri}.
The possible existence of a fourth fermion generation is also considered in the
latest Particle Data Group Report \cite{pdg}, where mass bounds are listed.
For a recent model-independet analysis, see Ref.~\cite{nov}.

It is advantageous to trace such novel fermions via their loop effects in the
Higgs sector, since these effects are also sensitive to mass-degenerate
isodoublets via fermion-mass power corrections.
This has originally been observed in Ref.~\cite{cha}
in connection with the $f\bar fH$, $W^+W^-H$, and $ZZH$ couplings.
Moreover, the $ggH$ coupling may serve as a device to detect
mass-degenerate isodoublets of ultraheavy quarks \cite{wil},
although power corrections do not occur here.
By contrast, in the gauge sector, power corrections only appear in connection
with isospin breaking \cite{ros}.
The influence of quarks is amplified relative to the one of leptons because
they come in triplicate.
The quark-induced corrections are greatly affected by QCD effects.
The two-loop ${\cal O}(\alpha_sG_FM_Q^2)$ corrections to
$\Gamma\left(H\to f\bar f\right)$ have recently been evaluated \cite{bak}.
In this paper, we shall extend that analysis to processes involving the
$W^+W^-H$ and $ZZH$ couplings.
The simplest processes of this kind are $H\to W^+W^-$ and $H\to ZZ$.
We shall also allow for one or both of the intermediate bosons to couple to
light-fermion currents.
Specifically, we shall consider $e^+e^-\to f\bar fH$ via Higgs-strahlung,
via $W^+W^-$ fusion (with $f=\nu_e$), and via $ZZ$ fusion (with $f=e$),
as well as the decays of the Higgs boson into four fermions via two
intermediate bosons.
In the case of five-point processes, we shall neglect interference terms
with a single fermion trace, since these are suppressed by
$\Gamma_V/M_V$, with $V=W,Z$.
For simplicity, we ignore the possibility of Cabibbo-Kobayashi-Maskawa mixings
between the external light fermions and the virtual high-mass quarks.

The hardest technical difficulty that needs to be tackled here is to solve
the two-loop three-point integrals in connection with the $W^+W^-H$ and $ZZH$
vertex corrections in ${\cal O}(\alpha_sG_FM_Q^2)$.
Similarly to the analysis of the ${\cal O}(\alpha_sG_FM_t^2)$ corrections to
$\Gamma\left(H\to b\bar b\right)$ \cite{hbb},
$\Gamma\left(Z\to f\bar fH\right)$, and $\sigma(e^+e^-\to ZH)$ \cite{lep},
we may take advantage of a particular low-energy theorem \cite{ell,vai}.
Generally speaking, this theorem relates the amplitudes of two processes which
differ by the insertion of an external Higgs-boson line carrying zero
four-momentum.
It may be derived by observing the following two points:
\begin{enumerate}
\item The interactions of the Higgs boson with the massive particles in the SM
emerge from their mass terms by substituting $M_i\to M_i(1+H/v)$, where $M_i$
is the mass of the respective particle, $H$ is the Higgs field, and $v$ is the
Higgs vacuum expectation value.
\item A Higgs boson with zero four-momentum is represented by a constant field.
\end{enumerate}
This immediately implies that a zero-momentum Higgs boson may be attached
to an amplitude, ${\cal M}(A\to B)$, by carrying out the operation
\begin{equation}
\label{let}
\lim_{p_H\to0}{\cal M}(A\to B+H)={1\over v}\sum_i
{M_i\partial\over\partial M_i}{\cal M}(A\to B),
\end{equation}
where $i$ runs over all massive particles which are involved in the transition
$A\to B$.
This low-energy theorem comes with two caveats:
\begin{enumerate}
\item The differential operator in Eq.~(\ref{let}) does not act on the $M_i$
appearing in coupling constants, since this would generate tree-level vertices
involving the Higgs boson that do not exist in the SM.
\item Equation~(\ref{let}) must be formulated for bare quantities
if it is to be applied beyond the leading order.
\end{enumerate}

This paper is organized as follows.
In Section~2, we review the ${\cal O}(G_FM_Q^2)$ corrections to
$\Gamma\left(H\to f\bar f\right)$, $\Gamma(H\to W^+W^-)$, and
$\Gamma(H\to ZZ)$ and derive those to $\sigma\left(e^+e^-\to f\bar fH\right)$
and $\Gamma(H\to2V\to4f)$ by invoking the so-called improved Born approximation
(IBA) \cite{iba}.
In Section~3, we construct, by means of the low-energy theorem (\ref{let}),
a heavy-quark effective Lagrangian for the $W^+W^-H$ and $ZZH$ interactions
which accommodates the ${\cal O}(G_FM_Q^2)$ and ${\cal O}(\alpha_sG_FM_Q^2)$
corrections,
and apply it along with the IBA to the processes discussed in Section~2.
In Section~4, we numerically analyze the phenomenological consequences of our 
results.
Section~5 contains our conclusions.

\section{One-loop results}

In this section, we review the leading one-loop effects on processes involving
a $W^+W^-H$ or $ZZH$ coupling in the presence of a generic doublet of novel
flavours, $(U,D)$, with masses $M_U,M_D\gg M_Z,M_H$ and quantum-number
assignments as in the first three fermion generations.
For completeness, we also consider the implications for the $f\bar fH$ Yukawa
couplings, assuming that $f$ does not mix with $U$ or $D$.
Throughout this paper, we employ dimensional regularization with
$n=4-2\epsilon$ space-time dimensions and a 't~Hooft mass, $\mu$, to keep the
coupling constants dimensionless.
As usual, we take $\gamma_5$ to be anticommuting.
We work in the on-mass-shell renormalization scheme \cite{sir}, with $G_F$ as a
basic parameter and the definition $c_w^2=1-s_w^2=M_W^2/M_Z^2$.

First, we recall that, at one loop, an additional $(U,D)$ doublet contributes
to the deviation of the $\rho$ parameter from unity,
$\Delta\rho=1-1/\rho$, the amount \cite{ros}
\begin{equation}
\label{rho}
\Delta\rho_1={N_c\over2}G\left({M_U^2+M_D^2\over2}
-{M_U^2M_D^2\over M_U^2-M_D^2}\ln{M_U^2\over M_D^2}\right)\ge0,
\end{equation}
where $N_c=1$ (3) for leptons (quarks) and $G=\left(G_F/2\pi^2\sqrt2\right)$.
Here and in the following, the subscripts 1 and 2 mark ${\cal O}(G_FM_Q^2)$
and ${\cal O}(\alpha_sG_FM_Q^2)$ contributions, respectively.
Equation~(\ref{rho}) is valid for $M_U$ and $M_D$ arbitrary. 
It is well known that $\Delta\rho$ measures the isospin breaking in the 
fermion sector; $\Delta\rho_1$ vanishes for $M_U=M_D$.
By contrast, the corresponding shifts, $\delta$, in the tree-level couplings
of the Higgs boson to physical particles are not quenched for $M_U=M_D$.
They have been calculated, in the one-loop approximation, for
$\Gamma\left(H\to f\bar f\right)$ in Ref.~\cite{hff}, for
$\Gamma(H\to W^+W^-)$ in Refs.~\cite{daw,hww}, and for $\Gamma(H\to ZZ)$ in
Refs.~\cite{daw,hzz}.
Writing the corrections to these observables in the form $K=(1+\delta)^2$ and
considering the limit $M_U,M_D\gg M_Z,M_H$, one has \cite{cha}
\begin{eqnarray}
\label{ffh}
\delta_1^u&\n=\n&
{\Delta\rho_1\over2}+{N_c\over6}G(M_U^2+M_D^2)>0,\\
\label{vvh}
\delta_1^{WWH}&\n=\n&\delta_1^{ZZH}
={\Delta\rho_1\over2}-{N_c\over3}G(M_U^2+M_D^2)<0.
\end{eqnarray}
Equation~(\ref{ffh}) refers to $\Gamma\left(H\to f\bar f\right)$, where $f$
does not mix with $U$ or $D$, so that only the renormalizations of the
Higgs-boson wave function and vacuum expectation value contribute.
The superscript $u$ is to indicate that this is a {\it universal} correction,
which occurs as a building block in the renormalization of any Higgs-boson
production and decay process.
On the other hand, Eq.~(\ref{vvh}) also contains genuine vertex corrections.
The equality of $\delta_1^{WWH}$ and $\delta_1^{ZZH}$ is broken
by subleading one-loop terms, of ${\cal O}(G_FM_H^2)$.
We anticipate that it is also spoiled by the leading two-loop QCD corrections,
of ${\cal O}(\alpha_sG_FM_Q^2)$, to be calculated in Section~3, unless $U$ and 
$D$ are mass degenerate.

In order to describe the production of the Higgs boson in high-energy
colliding-beam experiments, we have to consider the Feynman diagrams which
emerge from the $W^+W^-H$ and $ZZH$ vertices by linking the intermediate-boson
legs to light-fermion lines.
Then, Eq.~(\ref{vvh}) must be complemented by the ${\cal O}(G_FM_Q^2)$
corrections which arise from the gauge-boson propagators;
the gauge-boson wave-function renormalizations, which appear in connection 
with $\Gamma(H\to W^+W^-)$ and $\Gamma(H\to ZZ)$, do not receive such
corrections.
This may be achieved by invoking the IBA.
The IBA provides a systematic and convenient method to incorporate the 
dominant corrections of fermionic origin to processes within the gauge sector
of the SM.
These are contained in $\Delta\rho$ and 
$\Delta\alpha=1-\alpha/\overline\alpha$, which parameterizes the running of the
fine-structure constant from its value, $\alpha$, defined in Thomson scattering
to its value, $\overline\alpha$, measured at the $Z$-boson scale.
The recipe is as follows.
Starting from the Born formula expressed in terms of $c_w$, $s_w$, and 
$\alpha$, one substitutes
\begin{equation}
\alpha\to\overline\alpha={\alpha\over1-\Delta\alpha},\qquad
c_w^2\to\overline c_w^2=1-\overline s_w^2=c_w^2(1-\Delta\rho).
\end{equation}
To eliminate $\overline\alpha$ in favour of $G_F$, one exploits the relation
\begin{equation}
{\sqrt2\over\pi}G_F={\overline\alpha\over\overline s_w^2M_W^2}
={\overline\alpha\over\overline c_w^2\overline s_w^2M_Z^2}(1-\Delta\rho),
\end{equation}
which correctly accounts for the leading fermionic corrections.

We shall first concentrate on the processes with a $ZZH$ coupling.
Combining specific knowledge of $\delta^{ZZH}$ with the IBA, we obtain the
correction factors for $\sigma\left(f\bar f\to ZH\right)$,
$\Gamma\left(Z\to f\bar fH\right)$, and $\Gamma\left(H\to f\bar fZ\right)$ in
the form \cite{lep,zffh}
\begin{eqnarray}
\label{kone}
K_1^{(f)}&\n=\n&{(1+\delta^{ZZH})^2\over1-\Delta\rho}\,
{\overline v_f^2+a_f^2\over v_f^2+a_f^2}\nonumber\\
&\n=\n&1+2\delta^{ZZH}+
\left(1-8c_w^2{Q_fv_f\over v_f^2+a_f^2}\right)\Delta\rho,
\end{eqnarray}
where $v_f=2I_f-4s_w^2Q_f$, $\overline v_f=2I_f-4\overline s_w^2Q_f$,
$a_f=2I_f$, $Q_f$ is the electric charge of $f$ in units of the positron
charge, $I_f$ is the third component of weak isospin of the left-handed
component of $f$, and we have omitted terms of ${\cal O}(G_F^2M_Q^4)$ in the
second line.
The corresponding Born formulas may be found in Refs.~\cite{eezh,zffh,riz},
respectively.
Furthermore, the correction factors for
$\sigma\left(f_1\bar f_1\to f_2\bar f_2H\right)$ (via $f_1\bar f_1$
annihilation) and $\Gamma\left(H\to f_1\bar f_1f_2\bar f_2\right)$ (via a
$ZZ$ intermediate state) read \cite{gro}
\begin{eqnarray}
\label{ktwo}
K_2^{(f_1f_2)}&\n=\n&
{(1+\delta^{ZZH})^2\over(1-\Delta\rho)^2}\,
{\overline v_{f_1}^2+a_{f_1}^2\over v_{f_1}^2+a_{f_1}^2}\,
{\overline v_{f_2}^2+a_{f_2}^2\over v_{f_2}^2+a_{f_2}^2}\nonumber\\
&\n=\n&1+2\delta^{ZZH}+
2\left[1-4c_w^2\left({Q_{f_1}v_{f_1}\over v_{f_1}^2+a_{f_1}^2}
+{Q_{f_2}v_{f_2}\over v_{f_2}^2+a_{f_2}^2}\right)\right]\Delta\rho.
\end{eqnarray}
The corresponding tree-level results are listed in Refs.~\cite{kle,hzgg},
respectively.
Here and in the following, we neglect interference terms of five-point
amplitudes with a single fermion trace, since these are strongly suppressed,
by $\Gamma_V/M_V$, with $V=W,Z$.
Such terms have recently been included in a tree-level calculation of
$\Gamma(H\to2V\to4f)$ for $M_H\ll M_W$ \cite{asa}.
The formulas become slightly more complicated if a fermion line runs from the
initial state to the final state, {\it e.g.}, in the case of $ZZ$ fusion.
In the latter case, the Born cross section may be evaluated from
\begin{equation}
\label{zfus}
\sigma(f_1f_2\to f_1f_2H)={G_F^3M_Z^8\over64\pi^3\sqrt2s^2}
\left[\left(v_{f_1}^2+a_{f_1}^2\right)\left(v_{f_2}^2+a_{f_2}^2\right)A
\pm4v_{f_1}a_{f_1}v_{f_2}a_{f_2}B\right],
\end{equation}
where
\begin{eqnarray}
\label{int}
A&\n=\n&\int_{M_H^2/s}^1da\,f(a),\nonumber\\
B&\n=\n&\int_{M_H^2/s}^1da\,g(a),
\end{eqnarray}
$\sqrt s$ is the centre-of-mass energy, $f(a)$ and $g(a)$ are listed in 
Eq.~(A.9) of Ref.~\cite{eezh}, and the plus/minus sign refers to an odd/even
number of antifermions in the initial state.
{\it E.g.}, $e^+e^-\to e^+e^-H$ requires the plus sign.
From the IBA it follows on that the correction factor for Eq.~(\ref{zfus}) is
given by
\begin{eqnarray}
\label{kthree}
K_3^{(f_1f_2)}&\n=\n&
{(1+\delta^{ZZH})^2\over(1-\Delta\rho)^2}\,
{\left(\overline v_{f_1}^2+a_{f_1}^2\right)
\left(\overline v_{f_2}^2+a_{f_2}^2\right)A
\pm4\overline v_{f_1}a_{f_1}\overline v_{f_2}a_{f_2}B\over
\left(v_{f_1}^2+a_{f_1}^2\right)\left(v_{f_2}^2+a_{f_2}^2\right)A
\pm4v_{f_1}a_{f_1}v_{f_2}a_{f_2}B}\nonumber\\
&\n=\n&1+2\delta^{ZZH}+
2\left[1-{4c_w^2\over1+r}\left({Q_{f_1}v_{f_1}\over v_{f_1}^2+a_{f_1}^2}
+{Q_{f_2}v_{f_2}\over v_{f_2}^2+a_{f_2}^2}\right)
\right.\nonumber\\&\n\n&{}-\left.
{2c_w^2\over1+1/r}\left({Q_{f_1}\over v_{f_1}}+{Q_{f_2}\over v_{f_2}}\right)
\right]\Delta\rho,
\end{eqnarray}
where
\begin{equation}
r={\pm4v_{f_1}a_{f_1}v_{f_2}a_{f_2}B\over
\left(v_{f_1}^2+a_{f_1}^2\right)\left(v_{f_2}^2+a_{f_2}^2\right)A}.
\end{equation}
In practice, one has $|r|\ll1$ (see Table~\ref{tab:x}), so that
Eq.~(\ref{ktwo}) is approximately recovered.

The processes which correspond to a $W^+W^-H$ vertex with one or both of
the $W$ bosons coupled to light-fermion currents do not receive additional
dominant fermionic corrections beyond the factor
\begin{equation}
\label{wwh}
K_{WWH}=(1+\delta^{WWH})^2,
\end{equation}
which already corrects $\Gamma(H\to W^+W^-)$.
This may be understood by observing that $G_F$ is {\it defined} through the
radiative correction to a four-fermion charged-current process, namely the
muon decay.
The tree-level formulas for $\sigma\left(f\bar{f^\prime}\to W^\pm H\right)$,
$\Gamma\left(H\to f\bar{f^\prime}W^\pm\right)$, and
$\Gamma\left(H\to f_1\bar{f_1^\prime}f_2^\prime\bar f_2\right)$ (with a
$W^+W^-$ intermediate state) may be found in Refs.~\cite{eic,riz,hzgg},
respectively.
Here and in the following, $f^\prime$ denotes the isopartner of $f$.
The lowest-order cross section of $f_1f_2\to f_1^\prime f_2^\prime H$ via
$W^+W^-$ fusion is described by Eqs.~(\ref{zfus}) and (\ref{int}), with
$v_f=a_f=\sqrt2$ and $M_Z$ replaced by $M_W$.

The aim of this paper is to complete the knowledge of the
${\cal O}(\alpha_sG_FM_Q^2)$ corrections to the Higgs-boson production and
decay rates.
In the remainder of this section, we shall collect the results which are 
already known.
In the case of $\Delta\rho$, we have \cite{djo}
\begin{equation}
\label{rhotwo}
\Delta\rho_2=-{N_c\over4}C_FaG
\left[{M_U^2+M_D^2\over2}+F(M_U^2,M_D^2)\right]\le0,
\end{equation}
where $N_c=3$, $C_F=(N_c^2-1)/(2N_c)=4/3$, $a=\alpha_s(\mu)/\pi$, and
\begin{equation}
\label{fud}
F(u,d)=(u-d)\di\left(1-{d\over u}\right)+{d\over u-d}\ln{u\over d}
\left[u-{3u^2+d^2\over2(u-d)}\ln{u\over d}\right].
\end{equation}
Note that $F(u,d)=F(d,u)$.
From Eq.~(\ref{fud}), we may read off the properties $F(u,u)=-u$ and
$F(u,0)=\zeta(2)u$.
For $M_U=M_t$ and $M_D=0$, Eq.~(\ref{rhotwo}) reproduces the well-known
${\cal O}(\alpha_sG_FM_t^2)$ result \cite{ver}.
For later use, we observe that
\begin{equation}
\label{diff}
\sum_{Q=U,D}{M_Q^2\partial\over\partial M_Q^2}\Delta\rho_{1,2}=
\Delta\rho_{1,2}.
\end{equation}
The QCD correction to Eq.~(\ref{ffh}) reads \cite{bak}
\begin{equation}
\label{ffhtwo}
\delta_2^u
={\Delta\rho_2\over2}-{N_c\over8}C_FaG(M_U^2+M_D^2)<0.
\end{equation}

In the next section, we shall derive the ${\cal O}(\alpha_sG_FM_Q^2)$
corrections to $\delta^{WWH}$ and $\delta^{ZZH}$ by means of the low-energy
theorem (\ref{let}).
The formalism developed in this section to find the ${\cal O}(G_FM_Q^2)$
corrections to the four- and five-point processes with a $ZZH$ coupling readily
carries over to ${\cal O}(\alpha_sG_FM_Q^2)$.
If the external fermions are leptons, we just need to include in
Eqs.~(\ref{kone}), (\ref{ktwo}), and (\ref{kthree}) the corresponding terms of
$\delta^{ZZH}$ and $\Delta\rho$.
Similarly, the four- and five-point processes with a $W^+W^-H$ coupling are
then simply corrected by $K_{WWH}$ given in Eq.~(\ref{wwh}), with the
${\cal O}(\alpha_sG_FM_Q^2)$ term included.

\section{Effective Lagrangian}

In the following, we shall proceed along the lines of Ref.~\cite{lep},
where the ${\cal O}(\alpha_sG_FM_t^2)$ correction to the $ZZH$ vertex was
found by means of the low-energy theorem (\ref{let}), assuming that $m_b=0$. 
We extend that analysis by keeping the quark masses arbitrary and by
considering also the $W^+W^-H$ coupling.
We shall explicitly work out the $W^+W^-H$ case, which is more involved
technically.
The $ZZH$ results will then be listed without derivation.

The starting point of our analysis is the amplitude characterizing the
propagation of an on-shell $W$ boson in the presence of quantum effects due
a doublet $(U,D)$ of high-mass quarks,
\begin{equation}
\label{ww}
{\cal M}(W\to W)=(M_W^0)^2-\left.\Pi_{WW}(q^2)\right|_{q^2=(M_W^0)^2},
\end{equation}
where $\Pi_{WW}(q^2)$ is the transverse $W$-boson self-energy, at
four-momentum $q$, written in terms of bare parameters.
Here and in the following, bare parameters are marked by the superscript 0.
In the $G_F$ representation, $\Pi_{WW}(q^2)$ is proportional to $(M_W^0)^2$,
which originates from the two $UDW$ gauge couplings.
Apart from this prefactor, we may put $q^2=0$ in Eq.~(\ref{ww}),
since we are working in the high-$M_Q$ approximation.
The low-energy theorem (\ref{let}) now tells us that we may attach a
zero-momentum Higgs boson to the $W\to W$ transition amplitude by carrying
out the operation
\begin{equation}
\lim_{p_H\to0}{\cal M}(W\to W+H)={1\over v^0}\left(
\sum_{Q=U,D}{M_Q^0\partial\over\partial M_Q^0}
+{M_W^0\partial\over\partial M_W^0}\right){\cal M}(W\to W),
\end{equation}
where we must treat the overall factor $(M_W^0)^2$ of $\Pi_{WW}(0)$
in Eq.~(\ref{ww}) as a constant.
This leads us to
\begin{equation}
\lim_{p_H\to0}{\cal M}(W\to W+H)={2(M_W^0)^2\over v^0}(1+E),
\end{equation}
with
\begin{equation}
\label{defe}
E=-\sum_{Q=U,D}{(M_Q^0)^2\partial\over\partial(M_Q^0)^2}\,
{\Pi_{WW}(0)\over (M_W^0)^2}.
\end{equation}

We are now in the position to write down the heavy-quark effective
$W^+W^-H$ interaction Lagrangian,
\begin{equation}
\label{bare}
{\cal L}_{WWH}=2(M_W^0)^2(W_\mu^+)^0(W^{-\mu})^0{H^0\over v^0}(1+E).
\end{equation}
Then, we have to carry out the renormalization procedure, {\it i.e.},
we have to split the bare parameters into renormalized ones and counterterms.
We fix the counterterms according to the on-shell scheme.
In the case of the $W$-boson mass and wave function, we have
\begin{eqnarray}
(M_W^0)^2&\n=\n&M_W^2+\delta M_W^2,\nonumber\\
(W_\mu^\pm)^0&\n=\n&(1+\delta Z_W)^{1/2}W_\mu^\pm,
\end{eqnarray}
with
\begin{eqnarray}
\delta M_W^2&\n=\n&\Pi_{WW}(0),\nonumber\\
\delta Z_W&\n=\n&-\Pi_{WW}^\prime(0),
\end{eqnarray}
where we have neglected $M_W$ against $M_Q$ in the loop amplitudes.
For dimensional reasons, $\delta Z_W$ does not receive corrections in
${\cal O}(G_FM_Q^2)$ and ${\cal O}(\alpha_sG_FM_Q^2)$.
Furthermore, we have \cite{bak}
\begin{equation}
{H^0\over v^0}=2^{1/4}G_F^{1/2}H(1+\delta^u),
\end{equation}
where the ${\cal O}(G_FM_Q^2)$ and ${\cal O}(\alpha_sG_FM_Q^2)$ terms of
$\delta^u$ are given in Eqs.~(\ref{ffh}) and (\ref{ffhtwo}), respectively.
Putting everything together, we obtain the renormalized version of
Eq.~(\ref{bare}),
\begin{equation}
\label{reno}
{\cal L}_{WWH}=2^{5/4}G_F^{1/2}M_W^2W_\mu^+W^{-\mu}H(1+\delta^{WWH}),
\end{equation}
with
\begin{equation}
\label{dwwh}
\delta^{WWH}=\delta^u+{\delta M_W^2\over M_W^2}+E.
\end{equation}
In order for $\delta^{WWH}$ to be finite through ${\cal O}(\alpha_sG_FM_Q^2)$,
we still need to renormalize the masses of the $U$ and $D$ quarks in the
${\cal O}(G_FM_Q^2)$ expressions for $\delta M_W^2/M_W^2$ and $E$, {\it i.e.},
we need to substitute
\begin{equation}
M_Q^0=M_Q+\delta M_Q,
\end{equation}
with \cite{tar}
\begin{equation}
\label{massct}
{\delta M_Q\over M_Q}=-{a\over4}C_F
\left({4\pi\mu^2\over M_Q^2}\right)^\epsilon
\Gamma(1+\epsilon){3-2\epsilon\over\epsilon(1-2\epsilon)},
\end{equation}
where $\Gamma$ is Euler's gamma function.

For convenience, we introduce the short-hand notations $q=M_Q^2$ and
$W=\delta M_W^2/M_W^2$.
Quantities with (without) the superscript 0 are written in terms of $M_Q^0$
($M_Q$).
First, we shall check our formalism in ${\cal O}(G_FM_Q^2)$.
We extract from Refs.~\cite{cha,hww} the ${\cal O}(G_FM_Q^2)$ amplitudes,
\begin{eqnarray}
\label{wone}
W_1&\n=\n&-\Delta\rho_1-{N_c\over2}G\sum_QM_Q^2
\left({4\pi\mu^2\over M_Q^2}\right)^\epsilon\Gamma(1+\epsilon)
\left[{1\over\epsilon}+{\cal O}(\epsilon)\right],\\
\label{eone}
E_1&\n=\n&\Delta\rho_1+{N_c\over2}G\sum_QM_Q^2
\left({4\pi\mu^2\over M_Q^2}\right)^\epsilon\Gamma(1+\epsilon)
\left[{1\over\epsilon}-1+{\cal O}(\epsilon)\right].
\end{eqnarray}
Substituting Eqs.~(\ref{ffh}), (\ref{wone}), and (\ref{eone}) into
Eq.~(\ref{dwwh}), we reproduce Eq.~(\ref{vvh}).
Furthermore, with the help of Eq.~(\ref{diff}), we immediately verify
Eq.~(\ref{defe}) at one loop, viz.
\begin{equation}
\label{xone}
E_1=-\sum{q\partial\over\partial q}W_1.
\end{equation}

Now, we shall proceed to two loops.
We have \cite{djo}
\begin{equation}
\label{wtwo}
W_2=-\Delta\rho_2+
{N_c\over8}C_FaG\sum_{Q=U,D}
M_Q^2\left({4\pi\mu^2\over M_Q^2}\right)^{2\epsilon}\Gamma^2(1+\epsilon)
\left[{3\over\epsilon^2}+{11\over2\epsilon}+{31\over4}+{\cal O}(\epsilon)
\right],
\end{equation}
where $\Delta\rho_2$ is given in Eq.~(\ref{rhotwo}).
Notice that Eq.~(\ref{wtwo}) already contains the contributions proportional
to $\delta M_Q$ which emerge from the renormalization of the quark masses in
Eq.~(\ref{wone}).
We wish to compute $E_2$.
According to Eq.~(\ref{defe}), we have
\begin{equation}
\label{xtwo}
E_2^0=-\sum{q\partial\over\partial q}W_2^0.
\end{equation}
Furthermore, we have
\begin{eqnarray}
\label{xthree}
W_2&\n=\n&W_2^0+\delta W_2,\nonumber\\
E_2&\n=\n&E_2^0+\delta E_2,
\end{eqnarray}
where the counterterms are obtained by scaling the one-loop results,
\begin{eqnarray}
\label{xfour}
\delta W_2&\n=\n&\sum{\delta q\over q}\,{q\partial\over\partial q}W_1,
\nonumber\\
\delta E_2&\n=\n&\sum{\delta q\over q}\,{q\partial\over\partial q}E_1.
\end{eqnarray}
Using Eqs.~(\ref{xone}), (\ref{xtwo}), (\ref{xthree}), and (\ref{xfour})
along with
\begin{equation}
{q\partial\over\partial q}\,{\delta q\over q}=-\epsilon{\delta q\over q},
\end{equation}
which may be gleaned from Eq.~(\ref{massct}), we find
\begin{equation}
\label{etwo}
E_2=-\sum\left({q\partial\over\partial q}W_2
+\epsilon{\delta q\over q}\,{q\partial\over\partial q}W_1\right).
\end{equation}
Obviously, knowledge of the ${\cal O}(\epsilon)$ term of $W_1$ is not
necessary for our purposes.
Inserting Eqs.~(\ref{massct}), (\ref{wone}), and (\ref{wtwo})
into Eq.~(\ref{etwo}) and employing Eq.~(\ref{diff}),
we obtain the desired two-loop three-point amplitude,
\begin{equation}
\label{etwofinal}
E_2=\Delta\rho_2-{3\over2}C_Fa\Delta\rho_1
-{N_c\over8}C_FaG\sum_{Q=U,D}
M_Q^2\left({4\pi\mu^2\over M_Q^2}\right)^{2\epsilon}\Gamma^2(1+\epsilon)
\left[{3\over\epsilon^2}+{11\over2\epsilon}-{5\over4}+{\cal O}(\epsilon)
\right],
\end{equation}
where $\Delta\rho_1$ and $\Delta\rho_2$ are given in Eqs.~(\ref{rho}) and
(\ref{rhotwo}), respectively.
The sum of Eqs.~(\ref{ffhtwo}), (\ref{wtwo}), and (\ref{etwofinal})
is devoid of ultraviolet divergences and reads
\begin{eqnarray}
\delta_2^{WWH}&\n=\n&
{\Delta\rho_2\over2}-{3\over2}C_Fa\Delta\rho_1
+N_cC_FaG(M_U^2+M_D^2)\nonumber\\
&\n=\n&{\Delta\rho_2\over2}-3C_Fa\delta_1^{WWH}>0,
\end{eqnarray}
where $\delta_1^{WWH}$ is given in Eq.~(\ref{vvh}).
This completes the derivation of the effective $W^+W^-H$ interaction Lagrangian
(\ref{reno}).
We observe that $\delta_2^{WWH}$ weakens the negative effect of
$\delta_1^{WWH}$.
In the cases of no $(M_U=M_D)$ and maximum $(M_U\gg M_D)$ isospin breaking,
we have
\begin{eqnarray}
\delta^{WWH}&\n=\n&-{2\over3}N_cGM_U^2(1-3C_Fa)
\nonumber\\
&\n\approx\n&-2GM_U^2(1-1.27324\,\alpha_s),\nonumber\\
\delta^{WWH}&\n=\n&-{5\over24}N_cGM_U^2\left\{1+{3\over5}C_Fa
\left[\zeta(2)-{9\over2}\right]\right\}\nonumber\\
&\n\approx\n&-{5\over8}GM_U^2(1-0.72704\,\alpha_s),
\end{eqnarray}
respectively.

The derivation of the effective $ZZH$ interaction Lagrangian proceeds in close
analogy to the $W^+W^-H$ case and leads to
\begin{equation}
{\cal L}_{ZZH}=2^{1/4}G_F^{1/2}M_Z^2Z_\mu Z^\mu H(1+\delta^{ZZH}),
\end{equation}
with
\begin{eqnarray}
\label{zzh}
\delta^{ZZH}&\n=\n&{\Delta\rho\over2}
+\left(\delta_1^{ZZH}-{\Delta\rho_1\over2}\right)(1-3C_Fa)\nonumber\\
&\n=\n&\delta^{WWH}+{3\over2}C_Fa\Delta\rho_1.
\end{eqnarray}
Again, we have $\delta_1^{ZZH}<0<\delta_2^{ZZH}$, {\it i.e}, the
${\cal O}(G_FM_Q^2)$ term is partly compensated by its QCD correction.
For $M_U=M_D$, $\delta^{ZZH}$ coincides with $\delta^{WWH}$.
For $M_U\gg M_D$, we recover the result of Ref.~\cite{lep},
\begin{eqnarray}
\delta^{ZZH}&\n=\n&-{5\over24}N_cGM_U^2\left\{1+3C_Fa
\left[{\zeta(2)\over5}-{3\over2}\right]\right\}\nonumber\\
&\n\approx\n&-{5\over8}GM_U^2(1-1.49098\,\alpha_s).
\end{eqnarray}
It is interesting to observe, that $\delta^u$ may be written in a form similar
to the first line of Eq.~(\ref{zzh}), namely
\begin{equation}
\delta^u={\Delta\rho\over2}
+\left(\delta_1^u-{\Delta\rho_1\over2}\right)
\left(1-{3\over4}C_Fa\right).
\end{equation}
As a corollary, we note that $\delta_{1,2}^u$, $\delta_{1,2}^{WWH}$, and
$\delta_{1,2}^{ZZH}$ also satisfy identities similar to Eq.~(\ref{diff}).

If the external fermions are leptons, then we may implement the
${\cal O}(\alpha_sG_FM_Q^2)$ corrections to the four- and five-point processes
considered in Section~2 by evaluating the $K$ factors in Eqs.~(\ref{kone}),
(\ref{ktwo}), (\ref{kthree}), and (\ref{wwh}) with the QCD-corrected
expressions for $\delta^{WWH}$, $\delta^{ZZH}$, and $\Delta\rho$.
In the case of external quarks, we also need to include the leading-order QCD
corrections to their couplings with the intermediate bosons, since these will
combine with the ${\cal O}(G_FM_Q^2)$ corrections to give additional
${\cal O}(\alpha_sG_FM_Q^2)$ terms.
In the case of jet production, we have to include an additional factor
$[1+(3C_Fa/4)]$ for each quark pair in the final state.
Our formalism is also applicable to Higgs-boson production via quark-pair
annihilation at hadron colliders and via intermediate-boson fusion at hadron
and $ep$ colliders.
Then, the pure QCD corrections to these processes may be conveniently
incorporated by using the appropriate hadronic structure functions \cite{han}.

If we set $M_U=M_t$ and $M_D=0$, our formulas may also be used to describe the
loop corrections induced by the top quark.
Then, however, special care must be exercised if there is beauty in the
external legs.
Specifically, if a $b\bar b$ pair is produced via a virtual $Z$ boson,
{\it e.g.}, by $Z\to b\bar bH$, $H\to b\bar bZ$, and $e^+e^-\to b\bar bH$,
then we must substitute $\overline v_b=2I_b-4\overline s_w^2Q_b/(1+\tau)$ in
the corresponding $K$ factor and include an overall factor $(1+\tau)^2$,
where \cite{fle}
\begin{equation}
\tau=-{G\over2}M_t^2\left[1-{3\over2}\zeta(2)C_Fa\right].
\end{equation}
Consequently, the relevant $K$ factors in Eqs.~(\ref{kone}) and (\ref{ktwo})
become
\begin{eqnarray}
K_1^{(b)}&\n=\n&\left(1+{3\over4}C_Fa\right)
(1+\delta^{ZZH})^2{(1+\tau)^2\over1-\Delta\rho}\,
{\overline v_b^2+a_b^2\over v_b^2+a_b^2}\nonumber\\
&\n=\n&\left(1+{3\over4}C_Fa\right)
\left[1+2\delta^{ZZH}+\left(1-8c_w^2{Q_bv_b\over v_b^2+a_b^2}\right)\Delta\rho
\right.\nonumber\\
&\n\n&+\left.
2\left(1+4s_w^2{Q_bv_b\over v_b^2+a_b^2}\right)\tau\right],\nonumber\\
K_2^{(\ell b)}&\n=\n&\left(1+{3\over4}C_Fa\right)
(1+\delta^{ZZH})^2{(1+\tau)^2\over(1-\Delta\rho)^2}\,
{\overline v_\ell^2+a_\ell^2\over v_\ell^2+a_\ell^2}\,
{\overline v_b^2+a_b^2\over v_b^2+a_b^2}\nonumber\\
&\n=\n&\left(1+{3\over4}C_Fa\right)
\left\{1+2\delta^{ZZH}
+2\left[1-4c_w^2\left({Q_\ell v_\ell\over v_\ell^2+a_\ell^2}
+{Q_bv_b\over v_b^2+a_b^2}\right)\right]\Delta\rho
\right.\nonumber\\
&\n\n&+\left.
2\left(1+4s_w^2{Q_bv_b\over v_b^2+a_b^2}\right)\tau\right\},
\end{eqnarray}
respectively.

\section{Numerical results}

We are now in a position to explore the phenomenological implications of our
results.
We take the values of our input parameters to be
$M_W=80.26$~GeV and $M_Z=91.1887$~GeV \cite{alv}, so that $s_w^2=0.2253$.

\begin{table}[ht]
\caption{Coefficients $C_1$ (upper entries) and $C_2$ (lower entries) in
Eq.~(\ref{kfac}) as functions of $M_D/M_U$ for the various Higgs-boson decay
rates and production cross sections discussed in the text.
In the last line, $x=B/A$, where $A$ and $B$ are given by Eq.~(\ref{int}),
and terms of ${\cal O}(x^2)$ have been neglected.
}\label{tab:k}
\medskip
\begin{tabular}{|c|c|c|c|c|c|c|} \hline\hline
$M_D/M_U$ &
$  0$ & $  0.2$ & $  0.4$ & $  0.6$ & $  0.8$ & $  1$ \\
\hline
$\Delta\rho$ &
$  1$ & $  0.772$ & $  0.462$ & $  0.211$ & $  0.053$ & $  0$ \\
&
$ -2.860$ & $ -2.407$ & $ -2.166$ & $ -2.056$ & $ -2.011$ & $ -2$ \\
\hline
$K_{ffH}$ &
$  7/3$ & $  2.158$ & $  2.009$ & $  2.024$ & $  2.240$ & $  8/3$ \\
&
$ -1.797$ & $ -1.503$ & $ -1.268$ & $ -1.110$ & $ -1.024$ & $ -1$ \\
\hline
$K_{WWH}$ &
$ -5/3$ & $ -2.002$ & $ -2.631$ & $ -3.416$ & $ -4.320$ & $ -16/3$ \\
&
$ -2.284$ & $ -3.072$ & $ -3.620$ & $ -3.873$ & $ -3.975$ & $ -4$ \\
\hline
$K_{ZZH}$ &
$ -5/3$ & $ -2.002$ & $ -2.631$ & $ -3.416$ & $ -4.320$ & $ -16/3$ \\
&
$ -4.684$ & $ -4.614$ & $ -4.322$ & $ -4.120$ & $ -4.024$ & $ -4$ \\
\hline
$K_1^{(\nu)}$ &
$ -2/3$ & $ -1.230$ & $ -2.170$ & $ -3.205$ & $ -4.267$ & $ -16/3$ \\
&
$ -7.420$ & $ -5.999$ & $ -4.781$ & $ -4.256$ & $ -4.050$ & $ -4$ \\
\hline
$K_1^{(\ell)}$ &
$ -1.272$ & $ -1.697$ & $ -2.449$ & $ -3.333$ & $ -4.299$ & $ -16/3$ \\
&
$ -5.249$ & $ -5.010$ & $ -4.482$ & $ -4.171$ & $ -4.034$ & $ -4$ \\
\hline
$K_1^{(u)}$ &
$ -2.089$ & $ -2.328$ & $ -2.827$ & $ -3.505$ & $ -4.343$ & $ -16/3$ \\
&
$ -4.315$ & $ -4.305$ & $ -4.173$ & $ -4.067$ & $ -4.014$ & $ -4$ \\
\hline
$K_1^{(d)}$ &
$ -1.637$ & $ -1.979$ & $ -2.618$ & $ -3.410$ & $ -4.319$ & $ -16/3$ \\
&
$ -4.717$ & $ -4.640$ & $ -4.333$ & $ -4.124$ & $ -4.025$ & $ -4$ \\
\hline
$K_2^{(\nu\nu)}$ &
$  1/3$ & $ -0.458$ & $ -1.708$ & $ -2.995$ & $ -4.214$ & $ -16/3$ \\
&
$  6.261$ & $-12.050$ & $ -5.488$ & $ -4.410$ & $ -4.075$ & $ -4$ \\
\hline
$K_2^{(\nu\ell)}$ &
$ -0.272$ & $ -0.925$ & $ -1.987$ & $ -3.122$ & $ -4.246$ & $ -16/3$ \\
&
$-14.025$ & $ -7.180$ & $ -5.021$ & $ -4.314$ & $ -4.060$ & $ -4$ \\
\hline
$K_2^{(\ell\ell)}$ &
$ -0.878$ & $ -1.393$ & $ -2.267$ & $ -3.250$ & $ -4.278$ & $ -16/3$ \\
&
$ -6.323$ & $ -5.579$ & $ -4.668$ & $ -4.225$ & $ -4.044$ & $ -4$ \\
\hline
$K_2^{(\ell u)}$ &
$ -1.695$ & $ -2.023$ & $ -2.644$ & $ -3.422$ & $ -4.322$ & $ -16/3$ \\
&
$ -4.654$ & $ -4.590$ & $ -4.311$ & $ -4.116$ & $ -4.024$ & $ -4$ \\
\hline
$K_2^{(\ell d)}$ &
$ -1.243$ & $ -1.674$ & $ -2.436$ & $ -3.327$ & $ -4.298$ & $ -16/3$ \\
&
$ -5.307$ & $ -5.045$ & $ -4.495$ & $ -4.175$ & $ -4.035$ & $ -4$ \\
\hline
$K_3^{(\ell\ell)}$ &
$( -0.878$ & $( -1.393$ & $( -2.267$ & $( -3.250$ & $( -4.278$ & $ -16/3$ \\
&
$ -2.353\,x)$ & $ -1.816\,x)$ & $ -1.087\,x)$ & $ -0.496\,x)$ & $ -0.125\,x)$
& \\
&
$( -6.323$ & $( -5.579$ & $( -4.668$ & $( -4.225$ & $( -4.044$ & $ -4$ \\
&
$+  9.281\,x)$ & $+  4.134\,x)$ & $+  1.200\,x)$ & $+  0.331\,x)$
& $+  0.059\,x)$ & \\
\hline\hline
\end{tabular}
\end{table}

In Eqs.~(\ref{kone}), (\ref{ktwo}), and (\ref{kthree}), we have presented
correction factors for various four- and five-point Higgs-boson production and
decay processes with a $ZZH$ coupling in terms of $\delta_{ZZH}$ and
$\Delta\rho$.
It is instructive to cast these correction factors into the generic form
\begin{equation}
\label{kfac}
K=1+{N_c\over4}GM_U^2C_1\left({M_D\over M_U}\right)\left[1
+{3\over4}C_FaC_2\left({M_D\over M_U}\right)\right],
\end{equation}
where $C_1$ and $C_2$ are dimensionless functions of $M_D/M_U$.
Since, in the high-$M_Q$ limit, $\Delta\rho$, $\delta^u$, $\delta^{WWH}$, and
$\delta^{ZZH}$ are symmetric in $M_U$ and $M_D$, we may, without loss of 
generality, assume that $M_D/M_U\le1$.
The specific forms of the prefactors are chosen in such a way that, in the case 
of the leading $M_t$-dependent contribution to $\Delta\rho$, the familiar 
values $C_1(0)=1$ \cite{ros} and $C_2(0)=(2/3)[2\zeta(2)+1]\approx2.860$
\cite{ver} are recovered.
Relative to $M_t=180$~GeV, we have
$(N_c/4)GM_U^2\approx1.015\%\times(M_U/M_t)^2$.
The outcome of this decomposition is displayed in Table~\ref{tab:k}, where
$C_1$ and $C_2$ are listed as functions of $M_D/M_U$ for various classes of 
processes with a $ZZH$ coupling.
Specifically, $K_1^{(f)}$ ($f=\nu,\ell,u,d$) refers to $Z\to f\bar fH$ and
$H\to f\bar fZ$,
$K_2^{(\nu\nu)}$ to $H\to\nu\bar\nu\nu^\prime\bar\nu^\prime$,
$K_2^{(\ell f)}$ to $e^+e^-\to f\bar fH$ via Higgs-strahlung,
and $K_3^{(\ell\ell)}$ to $e^+e^-\to e^+e^-H$ via $ZZ$ fusion.
For completeness, also $\Delta\rho$ and the correction factors $K_{ffH}$, 
$K_{WWH}$, and $K_{ZZH}$ for $\Gamma\left(H\to f\bar f\right)$,
$\Gamma(H\to W^+W^-)$, and $\Gamma(H\to ZZ)$, respectively, are considered.
As explained in Section~2, $K_{WWH}$ also applies to four- and five-point
processes with a $W^+W^-H$ coupling, such as $H\to f\bar{f^\prime}W^\pm$,
$H\to f_1\bar{f_1^\prime}f_2^\prime\bar f_2$, and
$e^+e^-\to \nu_e\bar\nu_eH$ via $W^+W^-$ fusion.
Notice that there are additional QCD corrections beyond Eq.~(\ref{kfac}) if 
external quarks are involved.
As discussed in Section~3, in the case of dijet production via an intermediate
boson, these give rise to an overall factor $[1+(3C_Fa/4)]$ on the right-hand
side of Eq.~(\ref{kfac}).
Such QCD corrections are not included in Table~\ref{tab:k}.

\begin{table}[ht]
\caption{Values of $-x$ in \% (upper entries) and $\sigma(e^+e^-\to e^+e^-H)$ 
in fb (lower entries) as functions of $M_H/s^{1/2}$ for selected values of
$s^{1/2}$.}
\label{tab:x}
\medskip
\begin{tabular}{|c|c|c|c|c|c|c|} \hline\hline
$M_H/\sqrt s$ & \multicolumn{6}{c|}{$\sqrt s$ [GeV]} \\
\cline{2-7}
& 175 & 300 & 500 & 1000 & 1500 & 2000 \\
\hline
0.3 & 13.623 &  5.103 &  1.603 &  0.271 &  0.088 &  0.039 \\
&  1.283 &  3.160 &  5.627 &  8.869 & 10.249 & 10.956 \\
\hline
0.4 & 14.417 &  5.550 &  1.754 &  0.292 &  0.094 &  0.040 \\
&  0.743 &  1.852 &  3.282 &  5.079 &  5.806 &  6.168 \\
\hline
0.5 & 15.490 &  6.210 &  1.997 &  0.332 &  0.106 &  0.045 \\
&  0.378 &  0.970 &  1.736 &  2.683 &  3.058 &  3.242 \\
\hline
0.6 & 16.946 &  7.206 &  2.392 &  0.403 &  0.128 &  0.055 \\
&  0.161 &  0.433 &  0.797 &  1.253 &  1.433 &  1.521 \\
\hline
0.7 & 18.958 &  8.786 &  3.083 &  0.537 &  0.173 &  0.074 \\
&  0.051 &  0.150 &  0.291 &  0.477 &  0.553 &  0.590 \\
\hline
0.8 & 21.837 & 11.535 &  4.473 &  0.838 &  0.276 &  0.120 \\
&  0.010 &  0.032 &  0.069 &  0.124 &  0.148 &  0.160 \\
\hline\hline
\end{tabular}
\end{table}

In the case of $K_3^{(\ell\ell)}$, we have treated $x=B/A$, where $A$ and $B$
are defined in Eq.~(\ref{int}), as an additional expansion parameter and
discarded terms of ${\cal O}(x^2)$.
This is justified because, in practice, $|x|\ll1$,
{\it e.g.}, for $\sqrt s=300$~GeV and $M_H=100$~GeV, we find
$x\approx-5.233\%$.
In Table~\ref{tab:x}, we list $-x$ (in \%) and $\sigma(e^+e^-\to e^+e^-H)$ (in 
fb) as functions of $M_H/\sqrt s$ for LEP2 energy and various envisaged NLC
energies.
We observe that $|x|$ decreases with $\sigma$ increasing and is at the few-\% 
level or below whenever $e^+e^-\to e^+e^-H$ is phenomenologically interesting.

Looking at Table~\ref{tab:k}, we see that, for all quantities except
$\Delta\rho$, $C_1$ grows in magnitude as $M_D/M_U$ approaches unity.
As is well known, $\Delta\rho$ is quenched in this limit. 
Moreover, the majority of the Higgs-related $K$ factors have $|C_1(0)|>1$,
{\it i.e.}, the corresponding observables are more sensitive to the existence
of novel fermion doublets than the $\rho$ parameter itself, even if isospin is
badly broken.
While $C_1>0$ for $K_{ffH}$, $C_1<0$ for all other Higgs-boson observables, 
with the exception of $K_2^{(\nu\nu)}$.
The case of $K_2^{(\nu\nu)}$ is special, since there $C_1$ changes sign, at
$M_D/M_U\approx0.113$.
Except for $K_2^{(\nu\nu)}$ with $M_D/M_U$ below this value, we always have
$C_2<0$, {\it i.e.}, the QCD corrections generally reduce the leading one-loop
terms in size.
In the presence of a $ZZH$ coupling, this screening effect is considerably
stronger than in the case of $\Delta\rho$.
In fact, for the $ZZH$-type processes, we throughout have
$C_2(M_D/M_U)\le C_2(1)=-4$.
Except in the small range $M_D/M_U\lsim0.095$, also $K_{WWH}$ exhibits a stronger
QCD screening than $\Delta\rho$.
$K_{WWH}$ and all $ZZH$-type $K$ factors coincide if $M_D/M_U=1$, since then 
$\delta^{WWH}=\delta^{ZZH}$ and $\Delta\rho=0$.

\section{Conclusions}

The implications of the possible existence of a fourth fermion generation for
electroweak physics have been extensively studied at one loop
\cite{kni,cel,cha,wil,ros,hff,daw,hww,hzz,eezh}.
Recently, this study has been extended to the two-loop level by analyzing the 
virtual QCD effects of ${\cal O}(\alpha_sG_FM_Q^2)$ due to a novel quark 
doublet, $(U,D)$, with arbitrary masses, in the gauge sector \cite{djo} and in
the $f\bar fH$ Yukawa couplings of the first three generations \cite{bak}.
In the present paper, this research program has been continued by 
investigating the ${\cal O}(\alpha_sG_FM_Q^2)$ corrections to the $W^+W^-H$ and 
$ZZH$ couplings.
In contrast to the vacuum-polarization analyses of Refs.~\cite{bak,djo}, this
involves two-loop three-point amplitudes, which are usually much harder to
compute.
To simplify matters, we assumed that $M_U$ and $M_D$ are large against the 
physical (invariant) masses of the on-shell (off-shell) $W$, $Z$, and Higgs 
bosons, which allowed us to take advantage of the low-energy theorem
(\ref{let}) \cite{ell,vai}.
The range of validity of this heavy-quark approximation may be defined more 
accurately by considering the thresholds in the relevant self-energy and 
vertex diagrams.
Then, it becomes apparent that the leading ${\cal O}(G_FM_Q^2)$ terms and 
their QCD corrections are expected to provide useful approximations to the 
full $M_Q$-dependent expressions as long as
$\min(M_U^2,M_D^2)\gg\max(p_{V_1}^2,p_{V_2}^2,p_H^2)/4$ is satisfied, where
$p_{V_1}$, $p_{V_2}$, and $p_H$ are the four-momenta flowing into the 
$V_1V_2H$ vertex of the considered process.
Assuming $M_D\le M_U$, this implies
$M_D\gg\sqrt s/2$ for Higgs-strahlung,
$M_D\gg\max\left(M_H,\sqrt{s-M_H^2}\right)/2$ for intermediate-boson fusion, and
$M_D\gg M_H/2$ for Higgs-boson decay.
In the case of Higgs-boson production, we have
$\sqrt s=M_Z$ for $Z\to f\bar fH$,
$\sqrt s>M_H$ for $e^+e^-\to f\bar fH$, and
$\sqrt s>M_Z+M_H$ for $e^+e^-\to ZH$,
while, in the case of Higgs-boson decay, we have
$M_H>0$ for $H\to4f$,
$M_H>M_V$ for $H\to V+2f$, and
$M_H>2M_V$ for $H\to2V$.
We recovered the notion, established in Refs.~\cite{bak,djo}, that, in the 
on-shell scheme implemented with $G_F$, the leading ${\cal O}(G_FM_Q^2)$ terms
get reduced in magnitude by their QCD corrections.
It turned out that, in general, this screening effect is considerably more
pronounced in the $W^+W^-H$ and $ZZH$ observables than in the electroweak
parameters \cite{djo} and Yukawa couplings \cite{bak}.
Nevertheless, the observables in the Higgs sector tend to be more sensitive to
the presence of novel fermion doublets, especially if isospin is only mildly
broken, in which case the $\rho$ parameter fails to serve as a useful probe.

In the discussion of fourth-generation scenarios, one has to bear in mind that
the novel-fermion masses must not exceed the vacuum-stability bound, which 
follows from the requirement that the running Higgs quartic coupling, 
$\lambda(\mu)$, must not turn negative for renormalization scales
$\mu<\Lambda$, where $\Lambda$ is the assumed mass scale of some new
interaction \cite{cab}.
This bound may be stringent, of order $v=246$~GeV, should there be a grand
dessert up to $\Lambda=\Lambda_{GUT}\approx10^{16}$~GeV, but it may be
considerably relaxed for $\Lambda$ in the few-TeV range.
Another theoretical difficulty related to $M_Q$ values in excess of $v$ is 
that the $Q\bar QH$ Yukawa coupling then becomes strong so that the 
Higgs-exchange corrections may not be negligible anymore.
Without explicit calculation, it is very difficult to predict above which 
values of $M_Q$ these corrections will surpass the QCD ones.
In the case of Higgs-boson production via gluon fusion, $gg\to H$, at the CERN
Large Hadron Collider, the QCD correction increases the lowest-order cross 
section by approximately 70\%, while, even for $M_Q=500$~GeV, the 
Higgs-related correction amounts to just 5\% \cite{gam}.
By analogy, this suggests that the two-loop Higgs-exchange contributions of
${\cal O}(G_F^2M_Q^4)$ to the $W^+W^-H$ and $ZZH$ observables are also likely
to be small as long as the vacuum-stability constraint is satisfied.
However, final clarity concerning this point can only come from a complete 
${\cal O}(G_F^2M_Q^4)$ calculation, which is a separate issue and lies beyond
the scope of the present work.

\bigskip
\centerline{\bf ACKNOWLEDGEMENTS}
\smallskip\noindent
The author would like to thank Michael Spira for useful discussions concerning 
the extension of the soft-Higgs theorem beyond the leading order.
He is indebted to the National Institute for Nuclear Theory in Seattle for its
great hospitality during a visit when this manuscript was finalized, and to the
Department of Energy for partial support during the completion of this work.


\begin{thebibliography}{99}

\bibitem{bjo} J.D. Bjorken, in {\it Weak Interactions at High Energy and the
Production of New Particles: Proceedings of Summer Institute on Particle
Physics}, August~2--13, 1976, edited by M.C. Zipf, SLAC Report No.~198 (1976)
p.~1.
%
\bibitem{jan} P. Janot, lecture delivered at {\it First General Meeting of the
LEP2 Workshop}, CERN, Geneva, Switzerland, 2--3 February 1995.
%
\bibitem{ell} J. Ellis, M.K. Gaillard, and D.V. Nanopoulos,
Nucl.\ Phys.\ {\bf B106}, 292 (1976).
%
\bibitem{iof} B.L. Ioffe and V.A. Khoze,
Fiz.\ Elem.\ Chastits At.\ Yadra {\bf9}, 118 (1978)
[Sov.\ J. Part.\ Nucl.\ {\bf9}, 50 (1978)].
%
\bibitem{kni} B.A. Kniehl, Phys.\ Rep.\ {\bf240}, 211 (1994).
%
\bibitem{abe} CDF Collaboration, F. Abe {\it et al.},
Phys.\ Rev.\ Lett.\ {\bf74}, 2626 (1995);
D0 Collaboration, S. Abachi {\it et al.},
Phys.\ Rev.\ Lett.\ {\bf74}, 2632 (1995).
%
\bibitem{hil} C.T. Hill and E.A. Paschos, Phys.\ Lett.\ B {\bf241}, 96 (1995).
%
\bibitem{lut} C.T. Hill, M.A. Luty, and E.A. Paschos,
Phys.\ Rev.\ D {\bf43}, 3011 (1991).
%
\bibitem{cab} N. Cabibbo, L. Maiani, G. Parisi, and R. Petronzio,
Nucl.\ Phys.\ {\bf B158}, 295 (1979).
%
\bibitem{ber} S. Bertolini and A. Sirlin, Phys.\ Lett.\ B {\bf257}, 179 (1991);
E. Gates and J. Terning, Phys.\ Rev.\ Lett.\ {\bf67}, 1840 (1991);
B.A. Kniehl and H.-G. Kohrs, Phys.\ Rev.\ D {\bf48}, 225 (1993).
%
\bibitem{cel} A. \c{C}elikel, A.K. \c{C}ift\c{c}i, and S. Sultansoy,
Phys.\ Lett.\ B {\bf342}, 257 (1995).
%
\bibitem{fri} H. Harari, H. Haut, and J. Weyers,
Phys.\ Lett.\ {\bf78B}, 459 (1978);
H. Fritzsch and J. Plankl, Phys.\ Lett.\ B {\bf237}, 451 (1990).
%
\bibitem{pdg} Particle Data Group, L. Montanet {\it et al.},
Phys.\ Rev.\ D {\bf50}, 1173 (1994).
%
\bibitem{nov} V.A. Novikov, L.B. Okun, A.N. Rozanov, M.I. Vysotsky, and
V.P. Yurov, Mod.\ Phys.\ Lett.\ A {\bf10}, 1915 (1995).
%
\bibitem{cha} M.S. Chanowitz, M.A. Furman, and I. Hinchliffe,
Phys.\ Lett.\ {\bf78B}, 285 (1978); Nucl.\ Phys.\ {\bf B153}, 402 (1979);
Z. Hioki, Phys.\ Lett.\ B {\bf224}, 417 (1989); {\bf228}, 560(E) (1989).
%
\bibitem{wil} F. Wilczek, Phys.\ Rev.\ Lett.\ {\bf39}, 1304 (1977).
%
\bibitem{ros} D.A. Ross and M. Veltman, Nucl.\ Phys.\ {\bf B95}, 135 (1975);
M. Veltman, Nucl.\ Phys.\ {\bf B123}, 89 (1977).
%
\bibitem{bak} B.A. Kniehl and A. Sirlin, Phys.\ Lett.\ B {\bf318}, 367 (1993);
B.A. Kniehl, Phys.\ Rev.\ D {\bf50}, 3314 (1994);
A. Djouadi and P. Gambino, Phys.\ Rev.\ D {\bf51}, 218 (1995).
%
\bibitem{hbb} B.A. Kniehl and M. Spira,
Nucl.\ Phys.\ {\bf B432}, 39 (1994).
%
\bibitem{lep} B.A. Kniehl and M. Spira,
Nucl.\ Phys.\ {\bf B443}, 37 (1995).
%
\bibitem{vai} A.I. Va\u\i nshte\u\i n, M.B. Voloshin, V.I. Zakharov, and
M.A. Shifman,
Yad.\ Fiz.\ {\bf30}, 1368 (1979) [Sov.\ J. Nucl.\ Phys.\ {\bf30}, 711 (1979)];
A.I. Va\u\i nshte\u\i n, V.I. Zakharov, and M.A. Shifman,
Usp.\ Fiz.\ Nauk {\bf131}, 537 (1980) [Sov.\ Phys.\ Usp.\ {\bf23}, 429 (1980)];
L.B. Okun, {\it Leptons and Quarks}, (North-Holland, Amsterdam, 1982) p.~229;
M.B. Voloshin, Yad.\ Fiz.\ {\bf44}, 738 (1986)
[Sov.\ J. Nucl.\ Phys.\ {\bf44}, 478 (1986)];
M.A. Shifman, Usp.\ Fiz.\ Nauk {\bf157}, 561 (1989)
[Sov.\ Phys.\ Usp.\ {\bf32}, 289 (1989)];
J.F. Gunion, H.E. Haber, G. Kane, and S. Dawson,
{\it The Higgs Hunter's Guide}, (Addison-Wesley, Redwood City, 1990) p.~40.
%
\bibitem{iba} M. Consoli, W. Hollik, and F. Jegerlehner,
in {\it $Z$ Physics at LEP~1}, edited by G. Altarelli, R. Kleiss, and
C. Verzegnassi, CERN Yellow Report No.\ 89--08 (1989) Vol.~1, p.~7;
W.F.L. Hollik, Fortschr.\ Phys.\ {\bf38}, 165 (1990);
F. Halzen, B.A. Kniehl, and M.L. Stong,
in {\it Particle Physics: VI Jorge Andr\'e Swieca Summer School},
Campos de Jord\~ao, Brasil, 14--26 January, 1991,
edited by O.J.P. \'Eboli, M. Gomes, and A. Santoro
(World Scientific, Singapore, 1992) p.~219;
Z. Phys.\ C {\bf58}, 119 (1993).
%
\bibitem{sir} A. Sirlin, Phys.\ Rev.\ D {\bf22}, 971 (1980);
K-I. Aoki, Z. Hioki, R. Kawabe, M. Konuma, and T. Muta,
Prog.\ Theor.\ Phys.\ Suppl.\ {\bf73}, 1 (1982);
M. B\"ohm, H. Spiesberger, and W. Hollik,
Fortschr.\ Phys.\ {\bf34}, 687 (1986).
%
\bibitem{hff} B.A. Kniehl, Nucl.\ Phys.\ {\bf B376}, 3 (1992).
%
\bibitem{daw} S. Dawson and S. Willenbrock,
Phys.\ Lett.\ B {\bf211}, 200 (1988).
%
\bibitem{hww} B.A. Kniehl, Nucl.\ Phys.\ {\bf B357}, 439 (1991).
%
\bibitem{hzz} B.A. Kniehl, Nucl.\ Phys.\ {\bf B352}, 1 (1991).
%
\bibitem{zffh} B.A. Kniehl, Phys.\ Lett.\ B {\bf282}, 249 (1992).
%
\bibitem{eezh} B.A. Kniehl, Z. Phys.\ C {\bf55}, 605 (1992).
%
\bibitem{riz} T.G. Rizzo, Phys.\ Rev.\ D {\bf22}, 722 (1980);
W.-Y. Keung and W.J. Marciano, Phys.\ Rev.\ D {\bf30}, 248 (1984).
%
\bibitem{gro} E. Gross, B.A. Kniehl, and G. Wolf,
Z. Phys.\ C {\bf63}, 417 (1994); {\bf66}, 321(E) (1995).
%
\bibitem{kle} F.A. Berends and R. Kleiss, Nucl.\ Phys.\ {\bf B260}, 32 (1985).
%
\bibitem{hzgg} B.A. Kniehl, Phys.\ Lett.\ B {\bf244}, 537 (1990);
A. Grau, G. Pancheri, and R.J.N. Phillips,
Phys.\ Lett.\ B {\bf251}, 293 (1990).
%
\bibitem{asa} T. Asaka and K. Hikasa, Phys.\ Lett.\ B {\bf345}, 36 (1995).
%
\bibitem{eic} E. Eichten, I. Hinchliffe, K. Lane, and C. Quigg,
Rev.\ Mod.\ Phys.\ {\bf56}, 579 (1984); {\bf58}, 1065(E) (1986).
%
\bibitem{djo} A. Djouadi and P. Gambino, Phys.\ Rev.\ D {\bf49}, 3499 (1994);
Phys.\ Rev.\ D {\bf49}, 4705 (1994).
%
\bibitem{ver} A. Djouadi and C. Verzegnassi,
Phys.\ Lett.\ B {\bf195}, 265 (1987);
A. Djouadi, Nuovo Cim.\ {\bf100A}, 357 (1988);
B.A. Kniehl, Nucl.\ Phys.\ {\bf B347}, 86 (1990).
%
\bibitem{tar} R. Tarrach, Nucl.\ Phys.\ {\bf B183}, 384 (1981).
%
\bibitem{han} T. Han and S. Willenbrock, Phys.\ Lett.\ B {\bf273}, 167 (1990);
T. Han, G. Valencia, and S. Willenbrock,
Phys.\ Rev.\ Lett.\ {\bf69}, 3274 (1992);
J. Bl\"umlein, G.J. van Oldenborgh, and R. R\"uckl,
Nucl.\ Phys.\ {\bf B395}, 35 (1993).
%
\bibitem{fle} J. Fleischer, O.V. Tarasov, F. Jegerlehner, and P. R\c aczka,
Phys.\ Lett.\ B {\bf293}, 437 (1992);
G. Buchalla and A.J. Buras, Nucl.\ Phys.\ {\bf B398}, 285 (1993);
G. Degrassi, Nucl.\ Phys.\ {\bf B407}, 271 (1993);
K.G. Chetyrkin, A. Kwiatkowski, and M. Steinhauser,
Mod.\ Phys.\ Lett.\ A {\bf8}, 2785 (1993).
%
\bibitem{alv} LEP Electroweak Working Group, M.G. Alviggi {\it et al.},
Report Nos.\ LEPEWWG/95--01, ALEPH~95--038 PHYSIC~95--036,
DELPHI~95--37 PHYS~482, L3 Note~1736, and OPAL Technical Note~TN284
(March 1995).
%
\bibitem{gam} A. Djouadi and P. Gambino,
Phys.\ Rev.\ Lett.\ {\bf73}, 2528 (1994).
%
\end{thebibliography}
\end{document}